\documentclass{JHEP3}

\usepackage{graphicx}    

\def\simgr{\mathrel{\hbox{\rlap{\hbox{\lower4pt\hbox{$\sim$}}}\hbox{$>$}}}}
\def\simls{\mathrel{\hbox{\rlap{\hbox{\lower4pt\hbox{$\sim$}}}\hbox{$<$}}}}

\title {On the quantum width of a black hole horizon}

\author{Donald Marolf\\
Physics Department, UCSB, Santa Barbara, CA 93106.
\texttt{marolf@physics.ucsb.edu}}

\abstract{The many low energy modes near a black hole
horizon give the thermal atmosphere a divergent entropy which
becomes of order $A/4G$ with a Planck scale cut-off. However,
Sorkin has given a Newtonian argument for 3+1 Schwarzschild black
holes to the effect that fluctuations of such modes provide the
horizon with a non-zero quantum mechanical width.  This width then
effectively enforces a cut-off at much larger distances so that
the entropy of the thermal atmosphere is negligible in comparison
with $A/4G$ for large black holes. We generalize and improve this
result by giving a relativistic argument valid for any spherical
black hole in any dimension.  The result is again a cut-off $L_c$
at a geometric mean of the Planck scale and the black hole radius;
in particular, $L_c^d \sim \frac{R}{T_H} \ell_p^{d-2}$. With this
cut-off, the entropy of the thermal atmosphere is again
parametrically small in comparison with the Bekenstein-Hawking
entropy of the black hole.  The effect of a large number $N$ of
fundamental fields and the discrepancies from naive predictions of
a stretched horizon model are also discussed.}

\date{December, 2003}

\begin{document}

\section{Introduction}

The work presented by the author at Adriatic 2003 comments on
arguments \cite{Bek,erice,LS,tHooft} which purport to derive
fundamental bounds on the entropy of any system from the so-called
generalized second law of thermodynamics; i.e. the second law
including the Bekenstein-Hawking entropy $S_{BH}= A/4G$ for black
holes. A number of interesting loopholes were pointed out stemming
from the effect of the thermal atmosphere (see \cite{MS1,MS2}) and
an additional ``observer dependence" in the concept of entropy
(see \cite{MMR}). However, this work is well described in
\cite{MS1,MS2} and \cite{MMR}.  Thus, rather than repeat the story
in detail, it seems best to use this proceedings to discuss other
thoughts on black hole entropy that are only vaguely related to
the material presented at the conference. Nonetheless, a brief
summary of that material will be included in the discussion below
where it ties in to the current storyline (section \ref{disc}).

The main focus here will be on the many low energy degrees of
freedom near the horizon of a black hole. Our primary interest
will be in non-degenerate (i.e., non-extremal) horizons, with the
degenerate case being treated only as a limit. Such degrees of
freedom give rise to an old puzzle reviewed in \cite{WaldRev}. The
puzzle is that, when described using quantum field theory in a
fixed curved background spacetime, such degrees of freedom give a
divergent contribution to the entropy of the black hole's thermal
atmosphere\footnote{i.e., the thermal bath of radiation constantly
being emitted and re-absorbed by the black hole due to the Hawking
effect.}.  If cut-off at the Planck scale, the thermal atmosphere
naturally yields \cite{atmS} an entropy of order the
Bekenstein-Hawking entropy $A/4G$ and potentially provides a large
correction to the entropy of the black hole. This is closely
related (see e.g. \cite{Muko} and \cite{RB}) to the suggestion
\cite{entangle} that the one should consider the entanglement
entropy of the quantum fields outside the black hole. In
\cite{atmS}, it was in fact suggested that this calculation might
in fact account for the entire entropy of the black hole, though
this idea appears to suffer from the so-called ``species problem"
and other issues discussed in \cite{WaldRev}.

Some time ago, it was argued by Sorkin \cite{wrinkle} that one
should cut-off the entropy of the thermal atmosphere at an even
larger distance from the black hole horizon. Sorkin used a
Newtonian model of a 3+1 Schwarzschild black hole to argue that
quantum fluctuations naturally cause the horizon to fluctuate on a
scale $L_c \sim (R\ell_p^2)^{1/3}$ where $R$ is the black hole
radius and $\ell_p$ is the Planck scale.  Since the fluctuations
are quantum in nature, we find it natural to describe this as
providing a quantum ``width" to the horizon.  Clearly then, one
cannot reliably characterize the region within $L_c$ of the
classical horizon location as ``outside" the black hole and it is
reasonable to suppose a cut-off at this scale on contributions to
the entropy of the (external) thermal atmosphere.  Quantum
fluctuations within $L_c$ of the classical horizon are then
perhaps better described as fluctuations of the black hole itself,
and may plausibly be assumed to already be included in the
Bekenstein-Hawking entropy of the black hole. (Though of course
the details of how or whether the full entropy is reflected in a
spacetime description remains unclear.)

Since the entropy of the atmosphere scales with $A/L_c^2$,
Sorkin's cut-off would make this contribution a parametrically
small correction to the Bekenstein-Hawking entropy of a large
black hole. However, one may ask to what extent Sorkin's Newtonian
model captures the relevant relativistic physics, and one may also
wonder whether a similar result is obtained for black holes with
charge or in different numbers of dimensions.  In particular,
Sorkin's Schwarzschild correction is large enough that, if the
same cut-off applied to the extremal case\footnote{Viewed as entropy of the thermal
atmosphere, one would expect the entropy correction to vanish as $T_H \rightarrow 0$, but
this is less clear when viewed as an entropy of entanglement across the horizon.}, one might expect it to
have been seen in stringy studies \cite{dW} of sub-leading
corrections to the Bekenstein-Hawking result for supersymmetric
black holes.

We report here on work to clarify these issues. A general
relativistic estimate is provided for the quantum width of the
horizon for arbitrary spherical black holes.  The result in $d$
spacetime dimensions is

\begin{equation}
L_c^d \sim \frac{\ell_p^{d-2} R N^{1/2}}{T_H},
\end{equation}
where $N$ is the number of effective free fields propagating near
the black hole and $T_H$ is the black hole's Hawking temperature.
Again we see that $A/L^{d-1}_c$ is parametrically small in
comparison with the Bekenstein-Hawking entropy, with the
additional feature that the correction also becomes parametrically
small\footnote{The divergence of $L_c$ is unphysical and signals a
breakdown of the near-horizon approximation used below, but
nevertheless $L_c$ is much larger for a low $T_H$ black hole than
for a black hole of similar size with $T_H \sim 1/R$.} as one
approaches extremality and $T_H \rightarrow 0$.  This is then in
agreement with the results of \cite{dW}, which finds no
contributions clearly associated with the thermal atmosphere of a
certain set of nearly extreme black holes.  The full correction
$NA/L_c^2$ can become large only when the number of fields $N$
becomes parametrically large.  It is amusing to note that for Schwarzschild black holes in 3+1
dimensions $L_c$ is only just below nuclear length scales for astrophysical black holes
and begins to approach atomic length scales for the largest known supermassive black holes.

These ideas and estimates are explained below.  We begin in
section \ref{modes} with a convenient description of the
near-horizon degrees of freedom.  We then perform the estimate in
section \ref{est}, which is based on a certain reasonable
conjecture with regard to distortions of the black hole horizon by
nearby objects.  We then discuss a few consequences and compare
with the natural expectation based on a ``stretched horizon" in
section \ref{disc}. Finally, we include an appendix which provides
an estimate which does not rely on the above-mentioned conjecture,
but which is clearly overly conservative. Nevertheless, this
conservative argument yields qualitatively similar results in $d <
6$ spacetime dimensions.

\section{Describing Near-horizon degrees of freedom}
\label{modes}

In flat spacetime, any mode of any field which occupies a small
region of space must have a correspondingly large energy. However,
the diverging redshift  at a horizon in principle allows for
localized degrees of freedom with significantly smaller Killing
energies.  A particularly well-known modern application of this
feature in the context of extremal horizons is the motivation
\cite{Juan} of the famous AdS/CFT conjecture.
It will be useful for our discussion to study the near-horizon
modes quantitatively.  We give a simple description below and refer the reader to, e.g., \cite{MTW} for a more
technical treatment using tortoise ($r^*$) coordinates.

Let us in particular consider a wavepacket
localized within a proper distance $L$ of the horizon, where $L$
is much smaller than the local curvature scale. We focus on the
non-extremal case so that, if the wavepacket is also localized on
a scale $L$ in the directions along the horizon, the situation is
well approximated by a wavepacket near a Rindler horizon in flat
space.

Low energy such wavepackets will have essentially no
variation on length scales shorter than $L$, except perhaps at
proper distances $s \ll L$ from the horizon where we are free to
take the wavepacket to vanish.  Thus, they will look much like the
larger wavepacket shown in figure \ref{fig:1} below.   Since the
energy density of the wavepacket is redshifted to zero near the
horizon, the dominant contribution to the energy of the wavepacket
will come from the part a proper distance $s \sim L$ away. Thus,
the total energy of the wavepacket is of order $(Lz)^{-1}$, where
$z$ is the redshift  at $s \sim L$ measured relative to infinity.
A simple way to obtain this redshift for arbitrary black holes is
to note that the local Unruh temperature at $s \sim L$ is of order
$1/L$, but it must also be related through the redshift to the
Hawking temperature $T_H$ of the black hole. Thus, $(Lz)^{-1} \sim
T_H$, and the wavepacket in question represents an excitation with
energy of order $T_H$.

\begin{figure}
\center
\includegraphics[height=3cm]{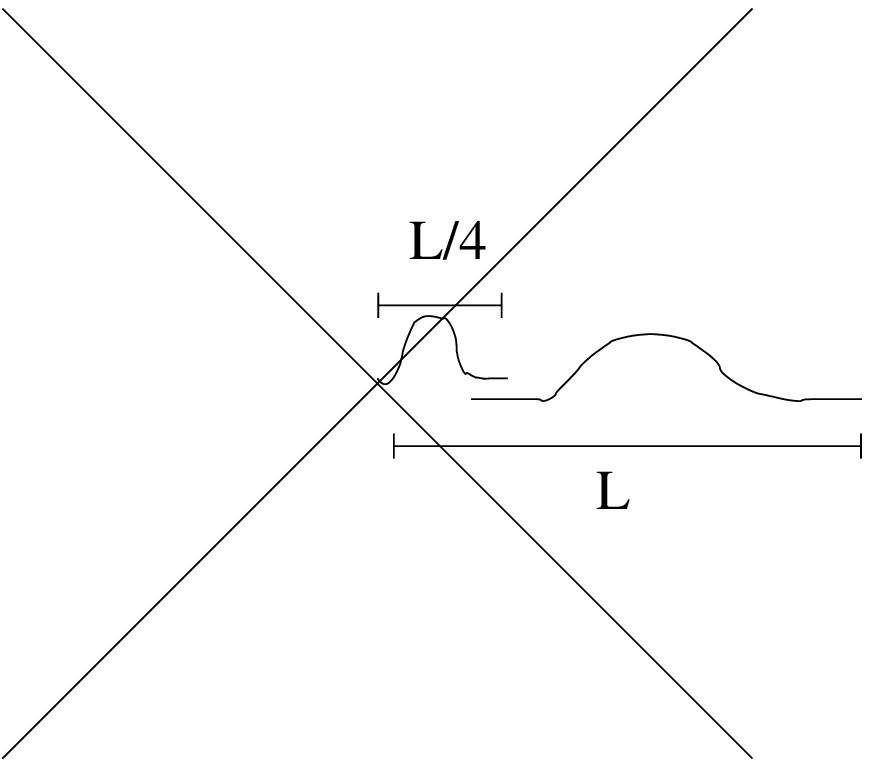}
\caption{Two orthogonal low energy  modes near a horizon. The larger has proper size $\sim L$ while
the smaller is $\sim L/4$.  Both have energy $\sim T_H$.}
\label{fig:1}       
\end{figure}

Note that we obtain such a wavepacket for each $L$, so that there are an infinite number of modes with energy $T_H$ near any
non-degenerate black hole horizon.  But the thermal atmosphere of the black hole is just this same system at temperature
$T_H$.  It is therefore clear that such modes contribute an infinite entropy.  In more detail, consider the number of such modes
contained between surfaces a distance $L$ and $L+dL$ from the horizon (for $L$ small compared to the curvature scale).
Since we considered wavepackets of size $L$ in each direction, it is useful to take each surface (which has area $A$) and  divide it into $A/L^2$ cells of size $L^2$.  

In fact, a standard calculation using tortoise ($r^*$) coordinates shows that
a careful choice of wavepacket shape can result in {\it arbitrarily} low energies, say $\epsilon T_H$ for $\epsilon \ll 1$, within
a proper distance $L$ of the horizon.  However, such modes cannot be localized as well on the sphere, so that there are
only $\lambda A/L^2$ of them and they contribute only ${\cal O}(1)$ changes in coefficients.

Finally, note that in the Rindler approximation a given mode is
transformed into the next mode closer to the horizon by scaling it
toward the bifurcation surface (see figure \ref{fig:1} again).  The
relevant measure in the radial direction is therefore scale
invariant and must be of order $dL/L$.  (This result 
also follows from the logarithmic relation between the tortoise coordinate ($r^*$)
and proper distance \cite{MTW}.) Thus, the total number of
modes between $L$ and $L+dL$ is of order $\frac{A}{L^3} dL$.
Integrating the total contribution down to some ultraviolet cutoff
$L_{UV}$ yields a total contribution of order $A/L^2_{UV}$. Each
mode with energy $\simls T_H$ contributes a few bits of entropy,
while modes with higher energy do not contribute. This then is the
origin of the observation \cite{atmS} that a Planck scale cutoff
yields an entropy of order the Bekenstein-Hawking entropy of the
black hole itself.  It can also be mapped (see e.g.
\cite{Muko,RB}) to the corresponding result \cite{entangle} for
entanglement entropy.

\section{Estimating the width}
\label{est}

The proper interpretation of the thermal atmosphere's entropy is
clearly an important  issue in black hole thermodynamics. For
example, in the most naive interpretation one might expect this to
be a next-order correction to the Bekenstein-Hawking entropy
$S_{BH}$.  Unfortunately, with a Planck scale cut-off it is of
comparable size to the ``zero order" contribution $S_{BH}$. One
may also try to cancel this entropy with a renormalization effect
(though this creates puzzles when various objects are lowered
toward the black hole horizon \cite{WaldRev}), or to interpret the
entropy of the thermal atmosphere as some sort of ``dual"
description of the Bekenstein-Hawking entropy itself
\cite{atmS,entangle}. Unfortunately, the latter approach suffers
from the well-known ``species problem" (i.e., it appears to depend
on the number of propagating fields near the black hole horizon)
and other concerns discussed in \cite{WaldRev}.

When the issue is phrased in terms of the near horizon modes,
Sorkin's potential resolution suggests itself  immediately. The
presence of an extra particle with fixed energy too close to the
horizon will effectively increase the mass of the black hole and
will cause the horizon to expand outward and engulf the particle.
If this particle is a quantum fluctuation associated
with Hawking radiation, then it is natural to describe the
effect as providing a non-zero quantum width for the horizon
itself.

Consider for simplicity a 3+1 Schwarzschild black hole and let $r$
be its Schwarzschild area-radius coordinate. The large entropy of
the thermal atmosphere comes from modes near the horizon with
energy $T_H$. But it is inconsistent to describe a mode with
energy $T_H$ supported below $r = R + 2G T_H$ in terms of quantum
field theory on a fixed spacetime containing a black hole of
radius $R$:  if we add a particle to such a mode, then considering
both the particle and the black hole we find a total energy $R/2G
+ T_H$ concentrated in a region smaller than the corresponding
Schwarzschild radius!  Thus,  if the particle's energy were
spherically distributed, one would expect that such a state is
better described by a larger black hole (of size  $R + 2G T_H$)
than by an excitation of the original black hole (of size $R$).
One therefore expects that fluctuations in the occupation numbers
of such modes are more properly described as excitations of the
horizon, and therefore that their contributions to the entropy are
already included in $S_{BH}$.  As a result, if one is considering
that part of the thermal atmosphere\footnote{This wording has been
used because the above observation is not necessarily in
disagreement with the idea that at least a part of the thermal
atmosphere's entropy is a dual description of $S_{BH}$.  On the
other hand, it is not obviously identical.  In particular, it does
not suffer from the issues discussed in \cite{WaldRev}.} which
acts as a {\it correction} to $S_{BH}$, then one should impose a
cutoff at some $r_c > R + 2G T_H$. Since $T_H \sim 1/R$ and since
for $r \ll R$ the proper distance $L$ from the horizon satisfies
$L^2/R \sim r -R$, one indeed arrives at the conclusion that the
cutoff must be at least a proper distance of order $L_c \sim
\ell_p$ from the horizon.

On the other hand, as discussed above there are {\it many} modes
near the horizon and, in calculating the entropy of the thermal
atmosphere we have considered {\it all} of these modes to be
thermally excited.  In particular, even for modes at a given scale
$L$ from the horizon, there are $R^2/L^2$ such modes due to the
size of the corresponding sphere around the black hole. Thus, a
more careful consideration is in order, and it is to this task
that we next turn. Note that it is necessarily {\it fluctuations}
of the occupations numbers that are relevant as the expectation
value of the stress-energy tensor remains small even very close to
the horizon, with the expected energy density being of order
$R^{-4}$.  Thus, a shell of thickness $\ell_p$ around the horizon
on average contains only an energy of order $(R^2  \ell_p) R^{-4}
= \ell_p R^{-2}$.  In contrast, we saw above that an energy of
order $T_H \sim R^{-1}$ is required to move the horizon out to
this radius. Thus, only departures from the mean can significantly
increase the cutoff beyond the naive Planck scale estimate.  Note also that energy fluctuations are largest
for modes with $E \sim T_H$.

As noted above, the effect of a spherical shell of mass on the
black hole horizon is easy to compute.  However, a localized
disturbance on a scale $L  \ll R$ is far from spherically
symmetric.  One could, of course, average over a sphere's worth of
fluctuations to estimate a typical fluctuation in the total mass
of the black hole.  Such an estimate is discussed in the appendix,
but is vastly over-conservative as the averaging will partially
cancel positive energy fluctuations on one side of the black hole
against negative energy fluctuations on the opposite side of the
black hole.  Thus, it will dramatically underestimate the {\it
local} fluctuations in the horizon location.  In other words, it
is clear that the largest effect comes from changes in the {\it
shape} of the horizon as opposed to just the overall {\it size}.

\begin{figure}
\center
\includegraphics[height=2cm]{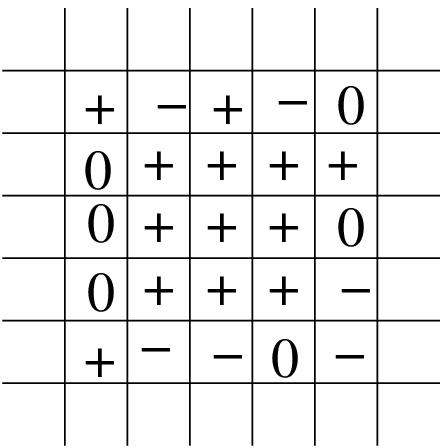}
\caption{The central fluctuation (+) happens to be surrounded by other fluctuations (+) of the same sign.  Fluctuations
of the opposite sign (-) may occur farther away, but the central fluctuation will not feel their influence on a
timescale
 $L$.}
\label{fig:2}       
\end{figure}

\begin{figure}
\center
\includegraphics[height=2cm]{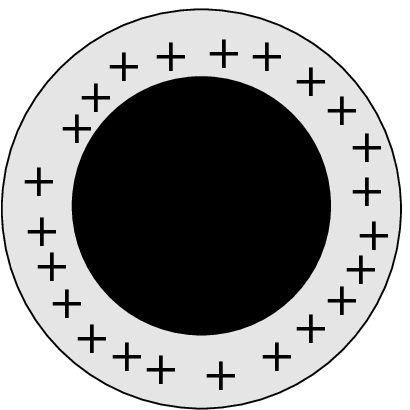}
\caption{A shell of fluctuations (+) around the black hole.  Since each fluctuation is of size $L$, a given fluctuation cannot detect the others on a timescale less than $L$.}
\label{fig:3}       
\end{figure}

How then shall we estimate this more localized effect on the
horizon? Consider a positive energy fluctuation of
length scale $L$ at a corresponding separation from the classical
horizon.  If this induces a bulge on the horizon which is
large enough to capture the fluctuation itself, then it is clear
that this must occur on a timescale\footnote{Say, as measured by
freely falling observers initially at rest with respect to the
black hole.  Since we are primarily concerned with the
perturbative regime, we may use the metric of the original
Black Hole to compute times to leading order.} $L$. Note that this
is also the natural lifetime of the fluctuation. Consider now the
center of the fluctuation. On a timescale $L$ the center can
receive no information from farther away than $L$. As a result, it
cannot know whether it is indeed part of a homogeneous spherical
shell of such fluctuations, or whether it is merely surrounded by
an additional layer or so of similar fluctuations\footnote{It may
just barely be able to tell whether the neighboring fluctuations
have the same sign, but such a clumping will occur a frequency
which is not parametrically small, and thus is large enough for
our purposes.} (see figure \ref{fig:2}). Thus, under reasonably
common conditions, we should get the right answer (as to whether
the horizon bulges outward and engulfs our fluctuation) by
supposing that the black hole is in fact surrounded by a spherical
shell of such fluctuations and determining whether this shell
would add enough mass to the black hole to enlarge the horizon
beyond the location of the fluctuations. Note that the shell has thickness $L$ (see figure \ref{fig:3}); luckily,
the calculation is just as easy for thick shells as for thin.

Let us consider a general spherically symmetric static metric of
the form

\begin{equation}
\label{Schw} ds^2 = - g_{tt}(r) dt^2 + g_{rr}(r) dr^2 + r^2
d\Omega^2_{d-2},
\end{equation}
where as usual $d\Omega^2_{d-2}$ is the metric on the unit
$(d-2)$-sphere.  We take $g_{tt}$ to have a first-order zero at $r=R$,
representing the non-degenerate black hole horizon.

In a sufficiently small region close to the horizon, we may
approximate the metric in the $r,t$ directions by the standard
Rindler metric:

\begin{equation}
\label{Rindler}
 ds^2 = - \kappa^2 \xi^2 dt^2 + d \xi^2 + R^2
d\Omega^2_{d-2} + higher \ order \ in \ \xi,
\end{equation}
where the constant $\kappa$ is the surface gravity, and is chosen
so that $t$ is properly normalized; e.g., so that $|\partial_t|
\rightarrow 1$ at infinity if the spacetime is asymptotically
flat. The Hawking temperature $T_H$ is then $T_H \sim \kappa$
since we set $\hbar =1$.  Note that for Schwarzschild black holes
the description (\ref{Rindler}) will be valid whenever $r \ll R$,
but for nearly extreme black holes we must approach the horizon
more closely than the difference in radius between the outer and
inner horizons.  We will always assume that the fluctuations are
close enough to the horizon for (\ref{Rindler}) to
apply\footnote{Any cut-off we find will be parametrically small in
the Planck length, so only for black holes parametrically
close to extremality can this approximation fail.}.

Let us ask how far the horizon moves outward when we add a mass
$\Delta m$ to the black hole but maintain spherical symmetry. The
new horizon is where $g_{tt}$ now vanishes.  Since it used to
vanish at $r=R$, we set
\begin{equation}
\label{d0}  0 =   \Delta g_{tt}\approx \partial_m g_{tt} \ \Delta
m \ +
\partial_r g_{tt} \ (r- R).
\end{equation}
On the other hand, comparing (\ref{Schw}) and (\ref{Rindler}), we
see that $g_{tt} \sim - T_H^2 \xi^2$, so we find
\begin{equation}
\label{obs}
\partial_r g_{tt} \sim -2 \xi T_H^2 \partial_r \xi = -2\xi T_H^2 \sqrt{g_{rr}}.
\end{equation}
Since $g_{tt}$ vanishes only to first order, we see that $g_{rr} =
\frac{c^2}{\xi^2} (1 + O(\xi)) $ for some constant $c$.  Thus  $r
- R \sim \xi^2/2c $.  Putting this together with (\ref{d0}) and
(\ref{obs}), one finds that adding a mass $\Delta m$ moves the
horizon to a new value of $r$ which, in the original metric, was
located at a proper distance
\begin{equation}
\label{xi}
 \xi  = \sqrt{ \frac{\partial_m g_{tt}}{T_H^2}  \Delta m
}
\end{equation}
from the original horizon.

Unfortunately, it seems that no general theorems concerning the
form of $\partial_m g_{tt}$ are available.  However, for the
Schwarzschild and Reissner-Nordstrom solutions in $d$ spacetime
dimensions one finds $\partial_m g_{tt} = (const) \ell^{d-2}_p
r^{-(d-3)}$ where the derivative is taken holding all charges
constant\footnote{This variation models
fluctuations in uncharged fields, such as the metric itself. One
could of course also consider fluctuations of charged fields.} and
$const$ is a number that depends at most on the dimension; i.e.,
which is independent of any charges or cosmological constant.  We
will therefore assume that any new black hole considered follows
this pattern.

With this understanding, one finds that the additional mass
required to, roughly speaking, ``move the horizon outward a proper
distance $\xi$" is given by
\begin{equation}
\label{sphere}
 \Delta m_{sphere} \sim \frac{ T_H^2 R^{d-3}\xi^2 }{   \ell_p^{d-2}} .
\end{equation}
The subscript $sphere$ has been added to remind us that we have
considered only spherically symmetric configurations above.  For
example, the outcome above might be attained by adding a spherical
shell of mass $\Delta m_{sphere}$ to our original black hole.

We now return to considering a single fluctuation of physical size
$L$ located at a proper distance $L$ from the classical horizon of
our black hole. We have seen that the typical energy of this
fluctuation is $\Delta m_{fluct} \sim T_H$, so that a spherical
shell of such fluctuations (as shown in figure \ref{fig:3}) would
have a mass $\Delta m_{sphere} \sim T_H (R/L)^{d-2}$. Comparison
with (\ref{sphere}) then shows that $L > \xi$ if and only if $L >
L_c$ where
\begin{equation}
L_c^{d} \sim \frac{R}{T_H} \ell_p^{d-2}.
\end{equation}
Putting together our logic above, this $L_c$ represents a rough
measure of the quantum width of the horizon and fluctuations
within $L_c$ of the classical horizon cannot be cleanly separated
from the black hole itself.

In general, there will be $N$ propagating degrees of freedom
(i.e., helicity states of quantum fields) at the scale $L_c$.
Any
combination of bosons and fermions leads to a typical increase in
the energy $\Delta m_{fluct}$ of fluctuations in a given region by a factor of
$\sqrt{N}$, so the cut-off length $L_c$ becomes
\begin{equation}
\label{cut} L_c^{d}\sim \frac{R}{T_H} N^{1/2}\ell_p^{d-2}.
\end{equation}
The entropy of the thermal atmosphere down to this cut-off is then
$S \sim NA/L_c^{d-2}$; i.e.,  smaller than $S_{BH} \sim
A/\ell_p^{d-2}$ unless $N$ is parametrically large in
$(R/\ell_p)$.  Note that $N$ should include any effective field
with a mass $m \simls 1/L$, so that in general $N$ depends on $L$
and (\ref{cut}) represents only an implicit solution for the
cut-off scale.

\section{Discussion}
\label{disc}

In the above sections we have argued that the horizon of any black
hole has an effective ``quantum width" given by (\ref{cut}). This
supports Sorkin's suggestion \cite{wrinkle} that fluctuations of
the horizon provide a cut-off on the entropy of the black hole's
thermal atmosphere, suppressing this entropy parametrically in
comparison with the black hole's Bekenstein-Hawking entropy. Our
work provides a fully relativistic treatment and includes both
uncharged black holes and those near the extremal limit, in
particular giving an explicit dependence of the cut-off on the
Hawking temperature separate from the dependence on the size $R$
of the black hole.  Our arguments concern general spherical black
holes with non-degenerate (i.e., non-extreme) horizons, but an
extension to the rotating case and a direct computation for
extreme black holes would clearly be of interest.
For Schwarzschild black holes in 3+1
dimensions, our estimate of the width $L_c$ is only just below nuclear length scales for astrophysical black holes
and begins to approach atomic length scales for the largest known supermassive black holes.

Our estimate of the width is based on the conjecture that whether
the horizon bulges outward to engulf a particular fluctuation
depends at most on whether the first few layers of surrounding
fluctuations have a similar sign (as in figure \ref{fig:2}), and
does not depend on the presence of fluctuations farther away. In
this case, we may model the calculation by assuming that there are
in fact a sphere's worth of such fluctuations and use simple
results for spherically symmetric perturbations of the black hole.
As discussed in the Appendix, a skeptic could raise some questions
about this conjecture, and a complete proof or counter-example
would be much desired.  The question should be amenable to study
by various perturbative methods, and we look forward to future
results in this direction\footnote{It would appear that the
closest result in the existing literature is provided by
\cite{AHs}, which considers two black holes of equal mass as
opposed to a single black hole with a small perturbation.}.

In the absence of such precise results, we may also take comfort
from the more conservative estimate given in the appendix.  This
estimate considers only the average fluctuation in the mass of a
spherical shell and so does not rely on any conjectures.  On the
other hand, it yields the desired result only for $d \le 5$.

Let us, however, proceed with the estimate (\ref{cut}).  One
interesting feature is that the entropy of the thermal atmosphere
continues to give a large contribution in the presence of a
sufficiently large number $N$ of propagating fields.  Under such
circumstances the black hole will rapidly decay into a fireball of
thermal radiation, unless the black hole is close to extremality
so that its temperature is low (in which case the thermal
atmosphere can be fully excited with only a small fraction of the
black hole's energy).

On a related note, we have seen that even for $N=1$ a large black
hole allows of order $(R/L_c)^{d-2}$ modes to be excited with an
energy of order $T_H$.  Consider now a particle placed in a mixed
state, which occupies an undetermined member of this set of modes.
The von-Neumann entropy of the corresponding density matrix is
then of order $\ln (R/L_c)$, which is parametrically larger than
$E/T_H \sim 1$. The reader might at first wonder what will happen
when such a particle falls into the black hole (since the black
hole's entropy will increase only by $E/T_H$ by the first law).

Indeed, following the logic of \cite{Bek}, one might expect it to
lead to a violation of the generalized second law (GSL). However,
as discussed in my talk at the conference (and as described in
\cite{MS1,MS2,MMR}) other, perhaps unexpected, effects will
intervene to save the GSL. Since the particle's energy is $T_H$,
it is clear that external observers will describe the particle as
being added to a rather busy thermal state, and not just to the
vacuum.  Such observers will be most concerned with the {\it
change} in entropy they assign to the process by which one
particle falls into the black hole but a thermal state remains
outside.  Note that since the thermal state is well-occupied, the
fundamental indistinguishability of particles may come into play
and one cannot guess the answer from a simple model of
distinguishable particles.  In fact, as discussed in \cite{MMR},
an ensemble of distinguishable particles which allows fluctuations
in particle number fails to be well-defined in our present
circumstance.

As follows from the corresponding analysis in \cite{MMR}, this
change in entropy does not in fact grow arbitrarily large with
$R/L$ but instead asymptotes to $E/T_H$, a value that preserves
the second law\footnote{A closely related phenomenon was
discovered in \cite{tele}, which observed that a background
of Unruh radiation downgrades the fidelity of quantum
teleportation, indicating a loss of access to quantum information
by observers who do not cross the horizon.}. It is only observers
who fall across the horizon which assign the particle an entropy
of order $\ln(R/L)$, but such observers do not see the particle
disappear and face no issues with regard to the GSL.

On the other hand, one can ask interesting questions about how the
experiences of such observers could be consistent with a
description of the black hole interior in terms of a Hilbert space
with a finite number $e^S_{BH}$ of dimensions.  We shall not
discuss such questions in detail here, as readers will no doubt
interpret the outcome according to their pre-existing views of
black hole entropy.  Some readers will see evidence that $S_{BH}$
counts only ``the ways the black hole can interact with the
external universe" and not the full set of interior states, while
others will see manifestations of ``complementarity" between
various observers inside the black hole.  The latter might be
protected by limitations on the transfer of information between
such observers analogous to those discussed in \cite{STU} for
observers comparing Hawking radiation with the original objects
falling into a black hole.

Returning now to somewhat firmer ground, perhaps the most
interesting observation regarding our result (\ref{cut}) is the
discrepancy with naive expectations based on a `stretched horizon'
picture of the black hole.  The stretched horizon is typically
described as a membrane just outside the black hole's classical
horizon having properties that make its classical dynamics
indistinguishable from the black hole itself \cite{MP}.  Various
differing proposals have been made for the details to incorporate
quantum effects; for example, \cite{STU,ST}  place the stretched
horizon on the symmetry sphere with $\ell_p^2$ more area than the
classical horizon while \cite{Sen} chooses the sphere with
string-scale temperature, and \cite{STU,KLL} choose the Planck
temperature sphere. However, at least for Schwarzschild black
holes, a membrane of Planck tension and Planck temperature
 (and thus located at roughly a
Planck scale proper distance from the classical horizon) naturally
reproduces the thermodynamics (e.g., energy and entropy) of the
black hole and so seems to be preferred. On the other hand, since
there are no infrared divergences in 2+1 or higher dimensions,
typical fluctuations in the location of such a membrane
(corresponding to black holes in $d \ge 4$) would also tend to be
of intrinsically Planck scale, and thus would be parametrically
smaller than our $L_c$. This suggests that the stretched horizon
concept may be in need of refinement; a process that may provide fertile ground for future research.

\medskip

{\bf Acknowledgments:} The author would like to thank  Raphael
Bousso, Ram Brustein, Chris Herzog, Gary Horowitz, Stefan Hollands, Dan Kabat, Luis Lehner, David
Lowe, Mark Spradlin, and the participants of the Adriatic 2003
conference for interesting discussions.  Special thanks go to
Djordje Minic, Simon Ross, and Rafael Sorkin for many discussions
during our collaborations, to Gary Horowitz and Eric Poisson
for help in locating reference \cite{AHs}, and to Luis Lehner for pointing out a number of typographic errors. 
This work was supported
in part by NSF grants PHY00-98747, PHY99-07949, and PHY03-42749,
and by funds from the University of California.

\appendix

\section{An overly conservative estimate}

In section \ref{est}, we argued that one could estimate local
distortions of the horizon caused by energy fluctuations by
considering a spherical shell of such fluctuations surrounding the
black hole.  The idea was that, due to the short timescales
involved, whether or not a given fluctuation was engulfed by a
distortion of the horizon should be independent of the existence
of neighboring fluctuations. However, the argument clearly falls
short of a rigorous proof and must therefore be termed a
conjecture.  In particular, a skeptic might worry that the
presence or absence of neighboring fluctuations over a time in the
past of order $R$ could affect the initial conditions in an
important way, invalidating the arguments of section \ref{est}.

Ultimately, this should be settled by a complete calculation.
However, not having such a calculation at hand, it is also of
interest to pursue other models of horizon fluctuations not based
on the conjecture above. To this end, let us now consider the
effect of averaging the fluctuations over a sphere around the
black hole. This is clearly overly conservative, as it averages
independent positive and negative fluctuations into an overall
expansion or contraction of the black hole horizon. However, it
provides a similar qualitative behavior to (\ref{cut}) in
spacetime dimensions $d \le 5$ and, in particular, reproduces
exactly Sorkin's Newtonian result in $d=4$.

The idea is a simple extension of our analysis so far.  We simply
note that there are of order $N(R/L)^{d-2}$ independent
fluctuations of size $L$ around the sphere.  As a result, a
typical fluctuation in the mass contained in this thick shell is
of order $\Delta m_{fluct} \sqrt{N (R/L)^{d-2} } = \frac{T_H
N^{1/2} R^{(d-2)/2}}{L^{(d-2)/2}}.$  Comparing this with
(\ref{xi}) yields
\begin{equation}
\label{weak}
 L_c^{(d+2)/2} \sim \frac{ N^{1/2} \ell_p^{d-2}}{T_H
R^{(d-4)/2}}.
\end{equation}

In particular, for a Schwarzschild black hole in $d=4$ one finds

\begin{equation}
L_c \simgr  N^{1/6} (R\ell_p^2)^{1/3},
\end{equation}
and again this cutoff leads to negligible entropy in the thermal
atmosphere.  However, since $T_H$ generally scales with $1/R$, we
see that (\ref{weak}) fails to suppress the thermal atmosphere's
entropy relative to the Bekenstein-Hawking entropy in $d \ge 6$
and that more care will be needed for such cases.

\end{document}